\documentclass[prb,aps,twocolumn,a4paper,showpacs,citeautoscript,floatfix]{revtex4}

\usepackage{amsmath,amssymb,amsfonts}
\usepackage{bm}\let\vec\bm
\usepackage{graphicx}

\begin{document}
\def\RH{R_{\text{H}}}

\title{Hall effect in strongly correlated low dimensional systems}

\author{G. Le\'{o}n}
\author{C. Berthod}
\author{T. Giamarchi}
\affiliation{DPMC-MaNEP, University of Geneva, 24 quai
Ernest-Ansermet, 1211 Geneva 4, Switzerland}

\begin{abstract}

We investigate the Hall effect in a quasi one-dimensional system made of weakly
coupled Luttinger liquids at half filling. Using a memory function approach, we
compute the Hall coefficient as a function of temperature and frequency in the
presence of umklapp scattering. We find a power-law correction to the
free-fermion value (band value), with an exponent depending on the Luttinger
parameter $K_{\rho}$. At sufficiently high temperature or frequency the Hall
coefficient approaches the band value.

\end{abstract}

\date{\today}
\maketitle

\section{Introduction}

The Hall effect has been continuously playing an important role in experimental
condensed-matter research, mostly because the interpretation of Hall
measurements is rather simple in classical Fermi systems \cite{ashcroft}. In
such materials the Hall coefficient is a remarkably robust property, which is
unaffected by interactions and only depends upon the shape of the Fermi surface
and the sign of the charge carriers. Deviations from this simple behavior are
generally taken as evidence for the onset of strong correlations and a failure
of the Fermi-liquid (FL) paradigm \cite{anderson_q1d}. Several authors have
investigated the Hall effect in threeand two-dimensional FL \cite{Shastry_Hall,
yakovenko_phenomenological_model_th, lange_hall_constant}, but the question of
the role of correlations in the Hall effect for low-dimensional systems remains
largely unexplored.

In most three-dimensional systems the interactions play a secondary role and the
FL picture is appropriate \cite{shankar_revue}. However, the prominence of
interactions increases as the dimensionality of the systems decreases and the FL
theory is believed to break down for many two-dimensional systems like,
\textit{e.g.}, the high-$T_c$ cuprate superconductors \cite{anderson_high_T_c}.
In one-dimensional (1D) systems interactions are dominant, and the FL
description must be replaced by the Luttinger liquid (LL) theory
\cite{Haldane_Luttinger,giamarchi_book_1d}. This theory predicts a rich variety
of physical phenomena, such as spin-charge separation or non-universal
temperature dependence of the transport properties
\cite{giamarchi_review_chemrev}, many of which have been observed
experimentally. Therefore large deviations from the classical Hall effect are
expected to occur in \emph{quasi}-one dimensional systems.

Among the various experimental realizations of low-dimensional systems (organic
conductors \cite{jerome_review_chemrev}, carbon nanotubes
\cite{dresselhaus_book_fullerenes_nanotubes}, ultra cold atomic gases
\cite{stoferle_tonks_optical}, etc.) the organic conductors are good
realizations of quasi-1D materials. Studies of the longitudinal transport have
successfully revealed signatures of LL properties
\cite{schwartz_electrodynamics, giamarchi_mott_shortrev, jerome_review_chemrev}.
Transport transverse to the chains has given access to the dimensional crossover
between a pure 1D behavior and a more conventional high-dimensional one
\cite{moser_conductivite_1d, henderson_transverse_optics_organics,
giamarchi_review_chemrev, jerome_review_chemrev}. To probe further the
consequences of correlations in these compounds, several groups have undertaken
the challenging measurement of the Hall coefficient $\RH(T)$
\cite{moser_hall_1d, mihaly_hall_1d, Korin-Hamzic_2003, Korin-Hamzic}. The
results, different depending on the direction of the applied magnetic field,
proved difficult to interpret due to a lack of theoretical understanding of this
problem. This prompted for a detailed theoretical analysis of the Hall effect in
quasi-1D systems. A first move in this direction was reported in
Ref.~~\onlinecite{lopatin_q1d_magnetooptical} where the Hall coefficient of
dissipationless weakly-coupled 1D interacting chains was computed and found to
be $T$-independent and equal to the band value. This surprising result shows
that in this case $\RH$, unlike other transport properties, is insensitive to
interactions. However the assumption of dissipationless chains is clearly too
crude to be compared with realistic systems for which a finite resistivity is
induced by the umklapp interactions \cite{giamarchi_umklapp_1d}.

In this work we examine the effect of umklapp scattering on the $T$-dependence
of the Hall coefficient in quasi-1D conductors. We consider $1/2$-filled 1D
chains and compute $\RH(T)$ to leading order in the umklapp scattering using the
memory function approach \cite{gotze_fonction_memoire,
proceedings_ISCOM06_Hall}. We find that the umklapp processes induce a
$T$-dependent correction to the free-fermions value, and this correction
decreases with increasing temperature as a power-law with an exponent depending
on interactions (Fig.~\ref{fig:graph}). We discuss the implications for quasi-1D
compounds.

\section{Model and method}

\begin{figure}[b]
\includegraphics[width=8.6cm]{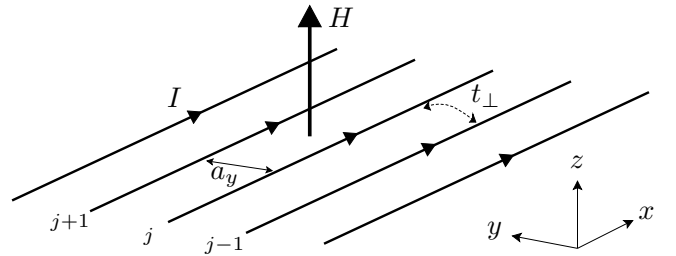}
\caption{\label{fig:model}
Schematics of the model. The chains and the current $I$ go along the $x$-axis,
the magnetic field $H$ is applied along the $z$-axis, and the Hall voltage is
measured along the $y$-axis.
}
\end{figure}

Our model is sketched in Fig.~\ref{fig:model}. We consider 1D chains coupled by
a hopping amplitude $t_{\perp}$ supposedly small compared to the in-chain
kinetic energy. The usual LL model of the 1D chains assumes that the electrons
have a linear dispersion with a velocity $v_{\text{F}}$. For a strictly linear
band, however, the Hall coefficient vanishes identically owing to particle-hole
symmetry. A band curvature close to the Fermi momenta $\pm k_{\text{F}}$ is thus
necessary to get a finite $\RH$. We therefore take for the 1D chains of
Fig.~\ref{fig:model} the dispersion
	\begin{equation}\label{dispersion}
		\xi_{\pm}(k)=\pm v_{\text{F}}(k\mp k_{\text{F}})
		+\alpha (k\mp k_{\text{F}})^2.
	\end{equation}
The upper (lower) sign corresponds to right (left) moving electrons. 
Eq.~(\ref{dispersion}) can be regarded as the minimal model which gives rise to
a Hall effect, while retaining most of the formal simplicity of the original LL
theory, and its wide domain of validity. In particular, this model is clearly
sufficient at low temperatures (compared to the electron bandwidth) since then
only electrons close to the Fermi points contribute to the conductivities.

Our purpose is to treat the umklapp term perturbatively. We express the
Hamiltonian as $\mathcal{H}_0+\mathcal{H}_u$ where $\mathcal{H}_u$ is the
umklapp scattering term and $\mathcal{H}_0$ reads
	\begin{multline}\label{Hamiltonian}
		\mathcal{H}_0=\int dx\sum_{j\sigma}\left[v_{\text{F}}
		\psi_{j\sigma}^{\dagger}\tau_3(-i\partial_x)
        \psi_{j\sigma}^{\phantom{\dagger}}
        -\alpha\,\psi_{j\sigma}^{\dagger}\partial_x^2
        \psi_{j\sigma}^{\phantom{\dagger}}\right.\\ \left.
        +g_2\,\psi_{j\sigma R}^{\dagger}
        \psi_{j\sigma R}^{\phantom{\dagger}}
        \psi_{j\sigma L}^{\dagger}
        \psi_{j\sigma L}^{\phantom{\dagger}}\right.\\ \left.
        -t_{\perp}\left(\psi_{j\sigma}^{\dagger}
        \psi_{j+1,\sigma}^{\phantom{\dagger}}
        e^{-ieA_{j,j+1}}+\text{h.c.}\right)\right].
	\end{multline}
In Eq.~(\ref{Hamiltonian}) $j$ is the chain index, $\tau_3$ is a Pauli matrix,
and $A_{j,j'}=\int_j^{j'}\vec{A}\cdot d\vec{l}$. We choose the Landau gauge
$A_y=Hx$, such that $A_{j,j+1}=Hxa_y$ with $a_y$ the interchain spacing.
$\psi^{\dagger}=(\psi^{\dagger}_R\;\psi^{\dagger}_L)$ is a two-component vector
composed of right- and left-moving electrons. The second term in
Eq.~(\ref{Hamiltonian}) is the band curvature, the third term is the forward
scattering and the last term corresponds to the coupling between the chains. In
Eq.~(\ref{Hamiltonian}) we have omitted the backscattering terms ($g_1$
processes) which are, for spin rotationally invariant systems, marginally
irrelevant \cite{giamarchi_book_1d}. We therefore take
$g_{1\perp}=g_{1\parallel}=0$. At $1/2$ filling the umklapp term reads
	\begin{equation}\label{Hu}
		\mathcal{H}_{u}=\frac{g_3}{2}\int dx\sum_{j\sigma}\left(
        \psi_{j\sigma R}^{\dagger}\psi_{j,-\sigma R}^{\dagger}
        \psi_{j\sigma L}^{\phantom{\dagger}}\psi_{j,-\sigma L}^{\phantom{\dagger}}
        +\text{h.c.}\right).
	\end{equation}
It corresponds to a process where two electrons with opposite spins change
direction by absorbing a momentum $4k_{\text{F}}$ from the lattice.

The Hall resistivity $\rho_{yx}$ relates to the conductivity tensor
$\sigma_{\mu\nu}$ through
	\begin{equation}
		\rho_{yx}=\frac{\sigma_{xy}}{\sigma_{xx}\sigma_{yy}+\sigma_{xy}^2}.
	\end{equation}
Here we calculate $\rho_{yx}$ using a memory function approach
\cite{gotze_fonction_memoire}. One rewrites the conductivity tensor in terms of
a memory matrix $\bm{M}(\omega)$ as
	\begin{equation}\label{sigma_M}
        \bm{\sigma}^T(\omega)=i\left\{\omega\openone+\bm{\chi}(0)
        \left[\bm{\Omega}+i\bm{M}(\omega)\right]\bm{\chi}^{-1}(0)
        \right\}^{-1}\bm{\chi}(0)
	\end{equation}
where $\bm{\sigma}^T$ denotes the transpose of $\bm{\sigma}$. The advantage
provided by the memory function is the possibility to make finite-order
perturbation expansions which are singular for the conductivities due to their
resonance structure \cite{gotze_fonction_memoire}. This formalism is especially
useful for LL (Ref.~~\onlinecite{giamarchi_umklapp_1d}) and was also used to
estimate the Hall coefficient in the 2D Hubbard model
\cite{lange_hall_constant}. $\bm{\chi}(0)$ is a diagonal matrix composed of the
diamagnetic susceptibilities in each direction,
$\bm{\chi}(0)=\left(\begin{smallmatrix}\chi_x(0)&0\\0&\chi_y(0)\end{smallmatrix}
\right)$, with
	\begin{equation}\label{chi0}
		\chi_{\mu}(0)=-\int dx\sum_j\left.\left\langle\frac{\partial^2\mathcal{H}}
		{\partial A_{\mu}^2(x)}\right\rangle_0\right|_{\vec{A}^{\text{el}}=0}.
	\end{equation}
The thermodynamic average $\langle\cdots\rangle_0$ is taken with respect to
$\mathcal{H}_0$ and $\mathcal{H}$ is the Hamiltonian of Eq.~(\ref{Hamiltonian})
in the presence of electric and magnetic fields,
$\vec{A}=\vec{A}^{\text{el}}+\vec{A}^{\text{mag}}$. The frequency matrix
$\mathbf{\Omega}$ in Eq.~(\ref{sigma_M}) is defined in terms of the equal-time
current-current correlators as \cite{lange_hall_constant}
	\begin{equation}\label{Omega}
		\Omega_{\mu\nu}=\frac{1}{\chi_{\mu}(0)}
		\left\langle\big[J_{\mu},J_{\nu}\big]\right\rangle.
	\end{equation}
From Eq.~(\ref{sigma_M}) one can directly express the memory matrix $\bm{M}$ in
terms of the conductivity tensor. In the following we will only need the
off-diagonal term $M_{xy}$ given by
	\begin{equation}\label{M}
		iM_{xy}(\omega)=\frac{i\chi_y(0)\sigma_{xy}(\omega)}
		{\sigma_{xx}(\omega)\sigma_{yy}(\omega)+\sigma_{xy}^2(\omega)}
		-\Omega_{xy}.
	\end{equation}
It is then straightforward to rewrite the Hall coefficient
$R_{\text{H}}=\rho_{yx}/H$ as
	\begin{equation}\label{RH}
		\RH(\omega)=\frac{1}{i\chi_y(0)}\lim_{H\to0}
		\frac{\Omega_{xy}+iM_{xy}(\omega)}{H}.
	\end{equation}

\section{Results}

From Eqs~(\ref{Hamiltonian}) and (\ref{chi0}) we obtain the longitudinal and
transverse diamagnetic terms as
	\begin{subequations}\begin{eqnarray}
        \label{chix}
        \chi_x(0)&=&-\frac{2e^2 v_{\text{F}}}{\pi a_y}\\
        \label{chiy}\nonumber
        \chi_y(0)&=&-2e^2t_{\perp}a_y^2\!\int dx
        \langle\psi_{0\uparrow}^{\dagger}(x)\psi_{1\uparrow}(x)
        e^{-ieHa_yx}+\text{h.c.}\rangle\\
    \end{eqnarray}\end{subequations}
For evaluating the frequency matrix we write down the current operators:
	\begin{subequations}\label{currents}\begin{eqnarray}
		\nonumber
        J_x&=&e\int dx\sum_{j\sigma}\psi_{j\sigma}^{\dagger}(x)
        \left[v_{\text{F}}\tau_3+2\alpha(-i\partial_x)\tau_1\right]
        \psi_{j\sigma}^{\phantom{\dagger}}(x)\\
        &&\\
        \nonumber
        J_y&=&-iet_{\perp}a_y\int dx\sum_{j\sigma}
        \left(\psi_{j\sigma}^{\dagger}\psi_{j+1,\sigma}^{\phantom{\dagger}}
        e^{-ieA_{j,j+1}}-\text{h.c.}\right)\\
    \end{eqnarray}\end{subequations}
The expression resulting from Eq.~(\ref{Omega}) for the frequency matrix is then
	\begin{equation}\label{Omegaxy}
		\Omega_{xy}=-i\frac{2\pi e\alpha t_{\perp}a_y^3H}{v_{\text{F}}}
        \!\int dx\,\langle\psi_{0\uparrow}^{\dagger}(x)
        \psi_{1\uparrow}^{\phantom{\dagger}}(x)
        e^{-ieHa_yx}+\text{h.c.}\rangle.
	\end{equation}
At this stage we can already evaluate the high-frequency limit of $\RH$, because
the memory matrix vanishes as $1/\omega^2$
(Refs~~\onlinecite{lange_hall_constant} and
~\onlinecite{gotze_fonction_memoire}) and thus $M_{xy}$ drops from
Eq.~(\ref{RH}) if $\omega\to\infty$. The effects of the umklapp disappear at
high frequency, and in this limit one recovers from
Eqs~(\ref{RH}--\ref{Omegaxy}) the result obtained for dissipationless chains
\cite{lopatin_q1d_magnetooptical}, namely that the Hall coefficient equals the
band value $\RH^0$:
	\begin{equation}\label{RH0}
		\RH(\infty)=\RH^0=\frac{\pi\alpha a_y}{ev_{\text{F}}}.
	\end{equation}
In the definition of the memory matrix, Eq.~(\ref{M}), we can ignore the terms
of order $H^2$ which do not contribute to $\RH$ in Eq.~(\ref{RH}). Furthermore
we express the conductivities in terms of current susceptibilities as
$\sigma_{\mu\nu}=\frac{i}{\omega}[\chi_{\mu\nu}-\delta_{\mu\nu}\chi_{\mu}(0)]$,
which leads to
	\begin{equation}
		iM_{xy}(\omega)=\frac{\omega\chi_y(0)\chi_{xy}(\omega)}
		{\left[\chi_x(0)-\chi_{xx}(\omega)\right]
		\left[\chi_y(0)-\chi_{yy}(\omega)\right]}-\Omega_{xy}.
	\end{equation}
We rewrite this formula at intermediate frequencies, such that
$|\chi_{\mu\mu}(\omega)/\chi_{\mu}(0)|$ is small. In this expansion we use the
equation of motion of the susceptibilitites \cite{gotze_fonction_memoire}, as
well as the relation
	$
		[\mathcal{H}_0,J_{\mu}]=-\Omega_{\nu\mu}J_{\nu}.
	$
For $\mu=x$ the latter
equation is exactly satisfied in our model, while for $\mu=y$
it is only verified in the continuum limit $a_y\to0$. The
resulting expression of the memory matrix is
	\begin{equation}\label{MofK}
		iM_{xy}(\omega)\approx-\frac{1}{\chi_x(0)}\frac{\langle K_x;K_y\rangle
		-\langle K_x;K_y\rangle_{\omega=0}}{\omega}
	\end{equation}
where $K_{\mu}$ are the \emph{residual forces} operators defined as the part of
the Hamiltonian which in the absence of magnetic field does not commute with the
currents, \textit{i.e.} $K_{\mu}=[\mathcal{H}_u,J_{\mu}]$, and $\langle
K_x;K_y\rangle$ stands for the retarded correlation function of the operators
$K_{\mu}$. The terms omitted in Eq.~(\ref{MofK}) are either of second order in
$|\chi_{\mu\mu}(\omega)/\chi_{\mu}(0)|$, of second order in $H$, or vanish in
the continuum limit $a_y\to0$. Using Eqs~(\ref{Hu}) and (\ref{currents}) we find
	\begin{subequations}\label{K}\begin{eqnarray}
		\nonumber
		K_x&=&2ev_{\text{F}}g_3\int dx\sum_{j\sigma}\\
		&&\left(\psi_{j\sigma R}^{\dagger}\psi_{j,-\sigma R}^{\dagger}
		\psi_{j,-\sigma L}\psi_{j\sigma L}-\text{h.c.}\right)\\
		\nonumber
		K_y&=&iet_{\perp}g_3a_y\int dx\sum_{j\sigma}\sum_{b=L,R}
		\Big[e^{-ieA_{j,j+1}}\\
		\nonumber
		&&\left(\psi_{j\sigma b}^{\dagger}\psi_{j,-\sigma b}^{\dagger}
		\psi_{j,-\sigma,-b}\psi_{j+1,\sigma,-b}\,-\right.\\
		&&\left.\psi_{j-1,\sigma b}^{\dagger}\psi_{j,-\sigma b}^{\dagger}
		\psi_{j,-\sigma,-b}\psi_{j\sigma,-b}\right)+\text{h.c.}\Big].\qquad
	\end{eqnarray}\end{subequations}
Note that each of the $K$'s is of first order in $g_3$, hence $M_{xy}$ is of
order $g_3^2$. The quantity $\langle K_x;K_y\rangle$ entering Eq.~(\ref{MofK})
is the real-frequency, long-wavelength limit of the correlator, which we
evaluate as
	\begin{multline}\label{KK}
		\langle K_x;K_y\rangle=-\int_0^{\beta} d\tau\,e^{i\Omega\tau}
		\langle T_{\tau}K_x(\tau)K_y(0)\rangle
		\Big|_{i\Omega\to\omega+i0^+}.
	\end{multline}
It is easy to prove that $\langle K_x;K_y\rangle$ vanishes for $H=0$ or
$\alpha=0$, by applying spatial inversion and particle-hole symmetry,
respectively. Retaining only leading-order terms in $t_{\perp}$ and $\alpha$,
the first nonvanishing contribution in Eq.~(\ref{KK}) is of order $\alpha
t_{\perp}^2g_3^2H$, and involves three spatial and three time integrations,
which we were not able to perform analytically. Based on a scaling analysis, we
can nevertheless extract the temperature (or frequency) dependence of this
contribution (see Appendix~\ref{app_scaling}), which yields:
	\begin{equation}\label{KK-T}
		\frac{1}{i\chi_x(0)\chi_y(0)}\frac{\langle K_x;K_y\rangle}{\omega H}
		\sim \alpha\,g_3^2\,\max(\omega,T)^{3K_{\rho}-3},
	\end{equation}
where $K_{\rho}$ is the LL parameter in the charge sector. In the absence of
interactions we have $K_{\rho}=1$, while $K_{\rho}<1$ ($K_{\rho}>1$) for
repulsive (attractive) interactions. If the interactions are strong and
repulsive ($K_\rho \ll 1$) the exponent in Eq.~(\ref{KK-T}) changes due to the
contraction \cite{giamarchi_book_1d} of the operators in $K_x$ and $K_y$, which
gives the relevant power-law in this case.

\begin{figure}[tb]
\includegraphics[width=7.5cm]{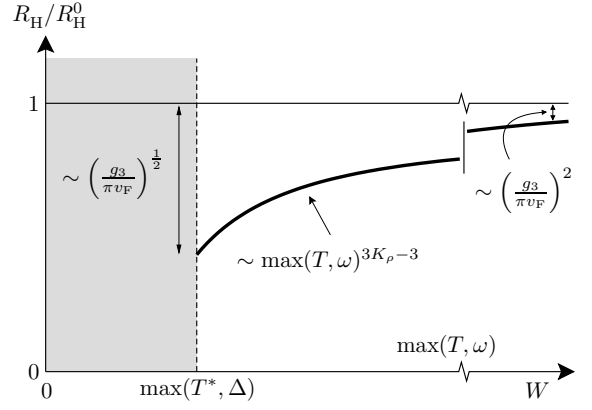}
\caption{\label{fig:graph}
Correction of the high-temperature/high-frequency Hall coefficient $\RH$ by the
umklapp scattering in weakly-coupled Luttinger liquids. $\RH^0$ is the value of
the Hall coefficient in the absence of umklapp scattering, Eq.~(\ref{RH0}), and
$W$ is the electron bandwidth. Our approach breaks down below some crossover
scale (dashed line, see text). In this figure we have assumed that $A$ in Eq.~(\ref{RH_result}) is negative.}
\end{figure}

Together with Eqs~(\ref{MofK}) and (\ref{RH}), Eq.~(\ref{KK-T}) leads to our
final expression for the Hall coefficient:
	\begin{equation}\label{RH_result}
		\RH=\RH^0\left[1+A\left(\frac{g_3}{\pi v_{\text{F}}}\right)^2
		\left(\frac{T}{W}\right)^{3K_{\rho}-3}\right]
	\end{equation}
with $W$ the electron bandwidth. Eq.~(\ref{RH_result}) shows that in
$1/2$-filled quasi-1D systems the umklapp scattering changes the absolute value
of the Hall coefficient with respect to the band value, which is only recovered
at high temperature or frequency. Note that Eq.~(\ref{RH_result}) also describes
the frequency dependence of $\RH$ provided $T$ is replaced by $\omega$. The
backscattering term $g_1$ (neglected here) could possibly give rise to
multiplicative logarithmic corrections to the power law in Eq.~(\ref{RH_result})
\cite{giamarchi_book_1d}. The sign of the dimensionless prefactor $A$ can only
be determined through a complete evaluation of $\langle K_x;K_y\rangle$ in
Eq.~(\ref{KK}), and is for the time being unknown. The available experimental
data are consistent with Eq.~(\ref{RH_result}) if one assumes that $A$ is
negative (see below), as illustrated in Fig.~\ref{fig:graph}.

\begin{figure}[tb]
\includegraphics[width=7.5cm]{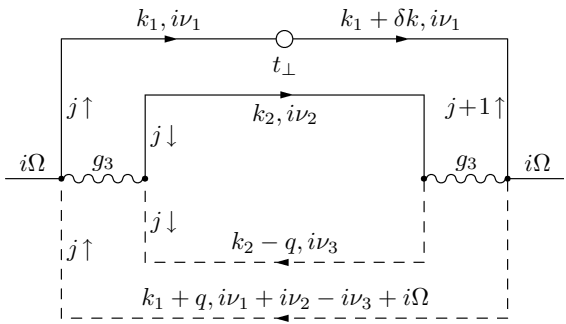}
\caption{\label{fig:diagram}
Example of a diagram appearing in Eq.~(\ref{KK}) at first order in $t_{\perp}$
and for $g_2=0$. The full (dashed) lines correspond to free right (left) moving
fermions, $j$ is the chain index, and the arrows represent up and down spins.
The magnetic field increases the momentum of the electron by $\delta k=eHa_y$.
}
\end{figure}

Eq.~(\ref{RH_result}) would imply that in the non-interacting limit
$K_{\rho}\to1$ ($g_2\to0$) the correction to the Hall coefficient behaves as
$\log(T/W)$. In order to check this prediction we have evaluated the correlator
in Eq.~(\ref{KK}) for $g_2=0$. The corresponding diagram sketched in
Fig.~\ref{fig:diagram} involves three frequency-momentum integrations, which in this case can
be fully worked out analytically (see Appendix~\ref{app_free}). The resulting
expression of $\RH$ at zero frequency is ($T\ll W$)
	\begin{equation}\label{no-int}
		\RH=\RH^0\left[1+\frac{1}{8}
		\left(\frac{g_3}{\pi v_{\text{F}}}\right)^2
        \log\left(\frac{T}{W}\right)\right],
	\end{equation}
consistently with Eq.~(\ref{RH_result}). For non-interacting electrons, though,
we see that the relative correction induced by the $1/2$-filling umklapp is
positive at $T<W$. Since all properties are analytic in $K_{\rho}$, we can also
deduce from Eqs~(\ref{RH_result}) and (\ref{no-int}) that $A$ tends to
$[24(1-K_{\rho})]^{-1}$ in the limit $K_{\rho}\to1$. Note that
Eq.~(\ref{no-int}) would also apply to models in which $g_2 \sim g_3$, such as
the Hubbard model, while Eq.~(\ref{RH_result}) is valid only when $g_3 \ll g_2$.

\section{Discussion and conclusion}

The result of Eq.~(\ref{RH_result}) shows that in $1/2$-filled quasi 1D systems the umklapp processes induce a correction to the free-fermion value (band value) of the Hall coefficient $\RH$, which depends on temperature as a power-law with an exponent depending on interactions. At high temperatures or frequencies, $\RH$ approaches the band value as shown in Fig.~\ref{fig:graph}, implying that any fitting of experimental data must be done with repect to the value of $\RH$ at high temperature or frequency.

To study the range of validity of our result, one must consider that at low 
temperature the quasi-1D systems generally enter either an insulating
state characterized by a Mott gap $\Delta$, or a coherent two- or
three-dimensional phase below a temperature $T^*$ controlled by
$t_{\perp}$;\cite{giamarchi_review_chemrev} in either case our model of
weakly-coupled LL is no longer valid, as illustrated in Fig.~\ref{fig:graph}.
The variations of $\RH$ below $\max(T^*,\Delta)$ can be very pronounced, and
depend strongly on the details of the materials. When the ground state is
insulating, for instance, $\RH(T)$ is expected to go through a minimum and
diverge like $e^{\Delta/T}$ as $T\to0$, reflecting the exponentially small
carrier density. Other behaviors, such as a change of sign due to the formation
of an ordered state or nesting in the FL regime
\cite{yakovenko_phenomenological_model_th}, can also occur. The validity of
Eq.~(\ref{RH_result}) is therefore limited to the LL domain:
$\max(T^*,\Delta)<\max(T,\omega)\ll W$.

For the case $\Delta>T^*$, we estimate the change of $\RH$ with respect to
$\RH^0$ at the crossover scale $\Delta$, for a system with $g_3\ll U$, where $U$
is the Coulomb repulsion. The umklapp-induced Mott gap in $1/2$-filled systems
is given \cite{giamarchi_book_1d} by $\Delta/W\sim[g_3/(\pi v_{\text{F}})]^x$
with $x=[2(1-K_{\rho})]^{-1}$. We thus find that the largest correction is
$\sim[g_3/(\pi v_{\text{F}})]^{\frac{1}{2}}$ and has a universal exponent. On
the other hand, $\RH$ approaches the asymptotic value $\RH^0$ quite slowly, and
according to Eq.~(\ref{RH_result}) a correction of $\sim[g_3/(\pi
v_{\text{F}})]^2$ still exists at temperatures comparable to the bandwidth.

The available Hall data in the TM family and in the geometry of the present
analysis \cite{moser_hall_1d,Korin-Hamzic_2003} show a weak correction to the free fermion value which depends on temperature. Some attempts to fit this behavior to a power law have been reported \cite{moser_hall_1d}. However the analysis was performed by fitting $\RH(T)$ to a power law starting at zero temperature. As explained above, the proper way to analyze the Hall effect in such quasi-1D systems is to fit the \emph{deviations} from the band value starting from the high temperature limit. It would be interesting to check whether a new analysis of the data would provide good agreement with our results. However in these compounds both $1/4$-filling and $1/2$-filling umklapp processes are present. 
For the longitudinal transport,
the $1/4$-filling contribution dominates \cite{giamarchi_review_chemrev}. For
the Hall effect, the analysis in the presence of $1/4$-filling umklapp is
considerably more involved, but a crude evaluation of the scaling properties of
the corresponding memory matrix gives also a weak power-law correction with an
exponent $2-16K_{\rho}+(K_{\rho}+K_{\rho}^{-1})/2$, and thus similar effects,
regardless of the dominant umklapp. The observed data is thus consistent with
the expected corrections coming from LL behavior. However more work, both
experimental and theoretical, is needed for the TM family because of this
additional complication, and to understand the data in a different geometry
where no temperature dependence is observed \cite{mihaly_hall_1d}.

Our result Eq.~(\ref{RH_result}) is however directly relevant for $1/2$-filled
organic conductors such as (TTM-TTP)I$_3$ and (DMTSA)BF$_4$
\cite{mori_review_chemrev}. Hall measurements for these compounds still remain
to be performed. Comparison of the Hall effect in these compounds with the one
in $1/4$-filled non-dimerized systems \cite{heuze_quarterfilled_refs,
jerome_review_chemrev} for which only $1/4$-filling umklapp is present, could
also help in understanding the dominant processes for the TM family.

\acknowledgments

This work was supported by the Swiss National Science Foundation through
Division II and MaNEP.

\appendix

\section{Scaling analysis}
\label{app_scaling}

In order to establish Eq.~(\ref{KK-T}) we evaluate the correlator $\langle
K_x;K_y\rangle$ to first order in $\alpha$ and $t_{\perp}$. Let's denote by
$\mathcal{H}_{\alpha}$ the curvature [second term in Eq.~(\ref{Hamiltonian})],
by $\mathcal{H}_{\perp}$ the inter-chain hopping [fourth term in
Eq.~(\ref{Hamiltonian})], and by $\mathcal{H}_{1D}$ the remaining part of the
Hamiltonian,
$\mathcal{H}_{1D}=\mathcal{H}_0-\mathcal{H}_{\alpha}-\mathcal{H}_{\perp}$.
Standard perturbation theory yields
	\begin{multline}\label{KKHcBC}
		\langle K_x;K_y\rangle=-\int d\tau\,e^{i\Omega\tau}
		\int d\tau_1d\tau_2\\
		\left\langle T_{\tau}K_x(\tau)K_y(0)\mathcal{H}_{\perp}(\tau_1)
		\mathcal{H}_{\alpha}(\tau_2)\right\rangle
	\end{multline}
where the average has to be taken with respect to $\mathcal{H}_{1D}$. The latter
corresponds to a 1D chain and can be easily bosonized,\cite{giamarchi_book_1d}
	\begin{eqnarray}\label{H_1D}
		\mathcal{H}_{1D}&=&\mathcal{H}_{\rho}+\mathcal{H}_{\sigma} \\
		\mathcal{H}_{\nu}&=&\int\frac{dx}{2\pi}\,\left\{u_{\nu}K_{\nu}
		\left[\nabla\theta_{\nu}(x)\right]^2+\frac{u_{\nu}}{K_{\nu}}
		\left[\nabla\phi_{\nu}(x)\right]^2\right\}\qquad
	\end{eqnarray}
where $\nu=\rho(\sigma)$ denotes the charge (spin) degrees of freedom, $u_{\nu}$
is a velocity, $K_{\nu}$ a dimensionless parameter, and $\theta_{\nu}$ and
$\phi_{\nu}$ are bosonic fields. In our case we have $K_{\rho}<1$ and
$K_{\sigma}=1$. The fields $\theta_{\nu}$ and $\phi_{\nu}$ and the fermionic
fields $\psi$ are related by
	\begin{equation}\label{fermionic}
		\psi_{\sigma b}(x)=\frac{e^{ibk_{\text{F}}x}}{\sqrt{2\pi a}}
		e^{-\frac{i}{\sqrt{2}}\left\{b\phi_{\rho}(x)-\theta_{\rho}(x)
		+\sigma\left[b\phi_{\sigma}(x)-\theta_{\sigma}(x)\right]\right\}}
	\end{equation}
with $b=+1(-1)$ corresponding to right (left) moving fermions, and $a$ a cutoff
of the order of the in-chain lattice spacing. With the help of
Eq.~(\ref{fermionic}) we bosonize each operator in Eq.~(\ref{KKHcBC}):
	\begin{eqnarray}
		K_x&=&\frac{4iev_{\text{F}}g_3}{(2\pi a)^2}\int dx\sum_{j\sigma}
			\sin\left(\sqrt{8}\phi_{\rho}(x)\right)_j\\
		\nonumber
		K_y&=&\frac{iet_{\perp}g_3a_y}{(2\pi a)^2}
			\sum_{\langle j,j'\rangle}\sum_{\sigma b}\int dx\,\epsilon_{jj'}
			\Big(e^{-ieHa_yx}\\
		\nonumber
		&&e^{\frac{i}{\sqrt{2}}\left\{3b\phi_{\rho}(x)-\epsilon_{jj'}\theta_{\rho}(x)
		-\sigma\left[b\phi_{\sigma}(x)+\epsilon_{jj'}\theta_{\sigma}(x)\right]\right\}_j}\\
		\nonumber
		&&e^{\frac{i}{\sqrt{2}}\left\{b\phi_{\rho}(x)+\epsilon_{jj'}\theta_{\rho}(x)
		+\sigma\left[b\phi_{\sigma}(x)+\epsilon_{jj'}\theta_{\sigma}(x)\right]\right\}_{j'}}\\
		&&+\text{h.c.}\Big)
	\end{eqnarray}
where $j$ and $j'$ are neighboring chains and $\epsilon_{jj'}=(-1)^{j'-j}$. For
the hopping term we have
	\begin{eqnarray}
		\nonumber
		\mathcal{H}_{\perp}&=&-\frac{t_{\perp}}{2\pi a}
		\sum_{j\sigma b}\int dx\Big(e^{-ieHa_yx}\\
		\nonumber
		&&e^{\frac{i}{\sqrt{2}}\left\{b\phi_{\rho}(x)-\theta_{\rho}(x)
		+\sigma\left[b\phi_{\sigma}(x)-\theta_{\sigma}(x)\right]\right\}_j}\\
		&&e^{-\frac{i}{\sqrt{2}}\left\{b\phi_{\rho}(x)-\theta_{\rho}(x)
		+\sigma\left[b\phi_{\sigma}(x)-\theta_{\sigma}(x)\right]\right\}_{j+1}}\\
		\nonumber
		&&+\text{h.c.}\Big)
	\end{eqnarray}
and for the band curvature term we take \cite{Haldane_JPC}
	\begin{equation}
		\mathcal{H}_{\alpha}=\frac{\alpha}{2\pi a} \int dx\,
		\frac{\left(\nabla\phi_{\rho}\right)^3}{2}.
	\end{equation}
Next we use the identity \cite{giamarchi_book_1d}
	\begin{multline}
		\langle\prod_ne^{i[A_n\phi(r_n)+B_n\theta(r_n)]}\rangle=
		\exp\Big\{-\frac{1}{2}{\sum_{n<m}}'\\
		-(A_nA_mK+B_nB_mK^{-1})F_1(r_n-r_m)\\
		+(A_nB_m+B_nA_m)F_2(r_n-r_m)\Big\}
	\end{multline}
where $r\equiv(x,u\tau)$, the notation ${\sum}'$ means that the sum is
restricted to those terms for which $\sum_nA_n=\sum_nB_n=0$, and $F_{1,2}$ are
universal functions. The resulting expression for the correlator in
Eq.~(\ref{KKHcBC}) is
	\begin{multline}
		\langle K_x;K_y\rangle\sim H\int d^2rd^2r_1d^2r_2\,
		e^{-3K_{\rho}F_1(r)}|r|\\
		e^{-K_{\rho}F_1(r-r_1)}
		e^{\frac{1}{2}(K_{\rho}-K_{\rho}^{-1}-2)F_1(r_1)}
		\frac{1}{|r_2|^3}.
	\end{multline}
The factor $|r|$ results from the linearization in $H$, and we have discarded
all factors involving the $F_2$ function, since they correspond to angular
integrals of the $r$ variables and therefore do not contribute to the scaling
dimension. At distances much larger than the cutoff $a$ we have
$e^{-AF_1(r)}\sim (a/|r|)^{A}$, and therefore we find the high temperature, high
frequency behavior as
	\begin{equation}
		\langle K_x;K_y\rangle\sim H
		\max(\omega,T)^{-3+4K_{\rho}-\frac{1}{2}(K_{\rho}-K_{\rho}^{-1})}.
	\end{equation}
We follow the same procedure for the diamagnetic term $\chi_y(0)$---however at
zeroth order in $\alpha$ and $H$---and find
	\begin{equation}
		\chi_y(0)\sim
		\max(\omega,T)^{-1+\frac{1}{2}(K_{\rho}+K_{\rho}^{-1})}.
	\end{equation}
Combining these expressions and collecting the relevant
prefactors we deduce Eq.~(\ref{KK-T}).

\section{Hall coefficient without forward scattering}
\label{app_free}

Here we provide the derivation of Eq.~(\ref{no-int}), which gives $\RH$ to
leading order in $g_3$ but in the absence of forward scattering ($g_2=0$). Using
Eqs~(\ref{RH}) and (\ref{RH0}) we can express the zero-frequency Hall
coefficient in terms of $\RH^0$ and $\text{Re}\,M_{xy}(i0^+)$. We then perform a
Kramers-Kronig transform, insert the free-fermion values of the diamagnetic
susceptibilities, $\chi_x(0)=-2e^2v_{\text{F}}/(\pi a_y)$ and
$\chi_y(0)=-4e^2t_{\perp}^2a_y/(\pi v_{\text{F}})$, and use Eq.~(\ref{MofK}) to
arrive at
	\begin{multline}\label{free_1}
		\RH(0)=\RH^0\left[1+\frac{v_{\text{F}}}{8e^3\alpha t_{\perp}^2a_y}
		\frac{1}{H}\int\frac{d\omega}{\omega^2}\right.\\
		\left.\text{Im}\left(i
		\left.\langle K_x;K_y\rangle_0\right|_{i\Omega\to\omega+i0^+}\right)\right]
	\end{multline}
where $\langle K_x;K_y\rangle_0$ is to be evaluated to first order in $H$. The
$\omega=0$ term in Eq.~(\ref{MofK}) disappears due to the principal part in the
$\omega$ integral in Eq.~(\ref{free_1}). From Eq.~(\ref{K}) one sees that
$\langle K_x;K_y\rangle_0$ involves 8 fermion fields and can be represented by
diagrams like the one displayed in Fig.~\ref{fig:diagram}. There are 32
different diagrams, but all of them can be expressed in terms of only one
function $A(i\Omega,H)$, whose expression is given by the diagram in
Fig.~\ref{fig:diagram}. We thus obtain
	\begin{multline}
		\RH(0)=\RH^0\left\{1-\frac{4v_{\text{F}}^2g_3^2}{e\alpha}
		\int\frac{d\omega}{\omega^2}\right.\\ \left.\phantom{\int}
		\text{Im}\left[A'(\omega+i0^+)-A(-\omega-i0^+)\right]\right\}
	\end{multline}
where $A'(i\Omega)=\partial A(i\Omega,H)/\partial H|_{H=0}$ and we have pulled
all prefactors from Eq.~(\ref{K}), as well as a factor $t_{\perp}$ from the
diagram, out of the definition of function $A$. The explicit expression of $A'$
is
\begin{widetext}
	\begin{multline}\label{eq:Aprime}
		A'(i\Omega)=\frac{e}{(2\pi)^3}\int dk_1dk_2dq\frac{d\,\xi_+(k_1)}{dk_1}
		\frac{1}{\beta^3}\sum_{\nu_1\nu_2\nu_3}\Big[\frac{1}{i\nu_1-\xi_+(k_1)}\Big]^3\\
		\frac{1}{i\nu_2-\xi_+(k_2)}\frac{1}{i\nu_3-\xi_-(k_2-q)}
		\frac{1}{i\nu_1+i\nu_2-i\nu_3+i\Omega-\xi_-(k_1+q)}.
	\end{multline}
\end{widetext}
The frequency summations in Eq.~(\ref{eq:Aprime}) are elementary, and the
various momentum integrals can also be evaluated analytically to first order in
$\alpha$, yielding
	\begin{multline}
		\RH(0)=\RH^0\Big[1-\frac{1}{16}\left(\frac{g_3}{\pi v_{\text{F}}}\right)^2
		\int\frac{d\omega}{\omega}\\
		\frac{(\beta\omega/4)^2-\sinh^2(\beta\omega/4)}
		{\tanh(\beta\omega/4)\sinh^2(\beta\omega/4)}\Big].
	\end{multline}
The remaining energy integral is divergent and must be regularized. Cutting the
integral at the bandwidth $W$ and assuming $T\ll W$ we obtain the asymptotic
behavior given in Eq.~(\ref{no-int}).

\end{document}